# Microwave Cytometry with Machine Learning for Shape-Resolved Microplastic Detection

Sayedus Salehin [1,2], Syed Shaheer Uddin Ahmed [1,2], Uzay Tefek [1,2], Laura Weirauch [3] and M. Selim Hanay [1,2,4*]

[1] Department of Mechanical Engineering, Bilkent University, 06800 Ankara, Turkey

[2] UNAM – Institute of Materials Science and Nanotechnology, Bilkent University, 06800 Ankara, Turkey

[3] Chemical Process Engineering, Faculty of Production Engineering, University of Bremen, 28359 Bremen, Germany

[4] Internation Iberian Nanotechnology Laboratory (INL), 4715-330, Braga, Portugal

**ABSTRACT:** Microplastics are increasingly recognized as a global environmental health threat, yet their detection and characterization remain constrained by the cost, form factor, and throughput of existing analytical tools. Portable micro/nanotechnology-based sensors are emerging to address this need, but most rely on the assumption of spherical particle geometry in their operating principle, limiting their relevance for environmental analysis. Here, we overcome this limitation by advancing microwave cytometry with machine learning–enabled shape recognition. Microwave cytometry is a flow-through electronic platform that integrates microwave resonator responses with low-frequency impedance signals to capture the dielectric signatures of individual particles. Using microscopy-derived shape measurements as ground truth, we trained a random forest model to decode these information-rich waveforms. Once trained, the system operates without optical input, enabling electronic-only determination of particle geometry. We demonstrate extraction of the major and minor axes of ellipsoidal microparticles with < 8% relative error on average and use these predictions to derive the dielectric permittivity of ellipsoid particles. This approach removes long-standing shape assumptions in microplastic sensing and establishes a pathway toward portable, high-throughput, morphology-aware detection technologies.

The ubiquitous presence of microplastics in the environment [1] and their associated impacts on the aquatic and terrestrial ecosystems [2] and the human health [3] underscores the importance of the continued efforts towards the development of effective microplastic detection and characterization methods. Fourier Transform Infrared Spectroscopy (FTIR) [4-6], Raman spectroscopy [7-9], and laser direct infrared (LDIR) spectroscopy [10-12] are characterization techniques being predominantly utilized by the scientific community for microplastic research. Although detection with LDIR is fast and can be automated, it is costly and requires significant amount of sample preparation [13]. While FTIR and Raman spectroscopy can detect microplastics with high accuracy, they are time consuming and requires expertise, and training [6, 7]. The characterization with Raman and FTIR for a single microplastic particle requires 10-15 minutes when operated by trained individuals [6]. In contrast, environmental monitoring requires reliable and rapid high throughput detection methods [2].

Due to the multifarious physiochemical processes that lead to its formation, microplastics are found in heterogenous shapes, and sizes in nature [1, 2, 14]. Shape of the microplastics is noted to be one of the most significant influencing factors for settling dynamics, atmospheric transport, deposition and fate of microplastics [15-18]. It also affects biological systems e.g., human body [19] as well as soil biota and organic matter decomposition [20, 21]. Shape of microplastics is commonly measured using stereo microscopy and optical microscopy prior to characterizing the microplastics with other measurement methods e.g., Raman spectroscopy, FTIR, in such cases the microplastics remain static [6]. Kim et al. have reported shape measurement of flowing ellipsoidal microplastics in a cross-slot microchannel using viscoelastic focusing with the help of optical microscope and high-speed camera [22]. It is interesting to note that, although shape of microplastics is critical to have environmental relevance, most publications on microplastics utilized spherical microplastics (63% of the studies used sphere and pellet shape) in their laboratory studies [23]. This consideration is even more critical in flow-through electronic sensors, where analytes are almost always restricted to commercially available spherical microparticles.

Machine learning (ML) has been used for characterizing micro and nano particles due to its ability to decipher the relationship between the experimental observations and the particle properties, especially from microscopy images. Some examples include nanorod size prediction from scanning electron microscopy images [24], tracking size and shape of synthesized gold nanoparticles by the combination of dynamic light scattering and UV-vis spectroscopy [25], the combined use of optical microscopy and impedance cytometry signals for pollen characterization [26] or to achieve an accuracy better than the cases where these techniques were used individually [27]. While an earlier work used extensional flow impedance cytometry to obtain electrical anisotropy indices for deformed red blood cell shape characterization [28], a more recent work has made use of ML models trained with impedance signal features from extensional flow impedance cytometry, and image derived shape metrics from image cytometry to obtain electrical shape anisotropy metrics for the quantification of cellular deformability across heterogenous cell populations [29]. ML has also been used to analyze the waveforms obtained in multichannel impedance cytometry [30].

Considering the need for portable and rapid microplastic detection, developing fully electronic sensing technologies capable of characterizing non-spherical microplastics is critical to advance our understanding of their environmental effect. To our knowledge, no studies were reported for the use of entirely electronic sensing in deciphering the shape properties of non-spherical microparticles. Here we developed such an approach by employing electronic sensing and optical microscopy only in the training phase and using only electronic sensing for the testing phase.

Recently, integrated microwave cytometry, a combination of high frequency capacitive sensor and a low frequency resistive sensor, has emerged as a powerful tool to rapidly characterize microparticles based on their permittivity. Utilizing microwave frequency, this technology overcomes the limitation of the impedance cytometry where at low frequency the ionic motion in liquid dominates the capacitive response of the material carrying the permittivity information. Prior work using this technology demonstrated the classification between polystyrene and micro-glass particles in the size range of <22 μm based on their permittivity with 94% accuracy [31]. By employing 3D metallized electrode in the sensing region, rapid differentiation between polystyrene and polyethylene microplastics in the 10-24 μm range was also reported [32].

In this work, we demonstrated the applicability of integrated microwave cytometry in detecting non-spherical microplastics and predict their shapes. Owing to the high permittivity contrast between the microplastic and the surrounding salt solution, individual microplastics traversing the sensing region induce high signal-to-noise-ratio spikes in the sensor signals and facilitates rapid detection. We calculated the Clausius-Mossotti factor of individual ellipsoid microplastics from experimental data and showed the deviation of this factor from that of spheres, this observation was supported by Monte Carlo simulations. Automated measurement of particles from microscopy videos using available image processing techniques can facilitate shape measurement of ellipsoidal microplastics for training the machine learning model since manual optical measurement would require a substantial amount of processing time and is prone to human error. Taking this into account, we developed an image processing workflow to automate the measurement of the shape of individual flowing microplastics from microscopy videos to train the machine learning model. Microscopic images are used only in the training phase as ground truths for microwave cytometry; the deployment phase does not require the usage of microscopy, and as such it is an electronics-only system. We demonstrated that machine learning algorithm can be utilized to predict the shape of individual non-spherical microplastics which unravels the relationship that exists between the features in electronic signal waveforms obtained from integrated microwave cytometry and the shape of the individual microplastics that cause those waveforms.

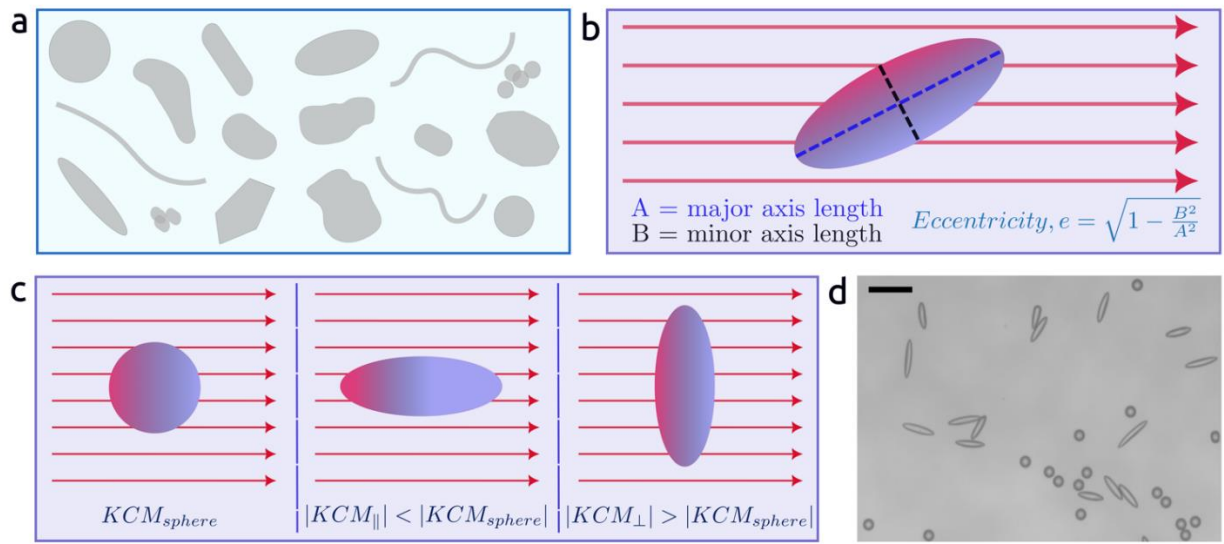

**Figure 1.** a) Representation of various shapes of microplastics as found in nature. b) Schematic of a non-spherical particle (prolate ellipsoidal particle) showing its major and minor axes, in an electric field represented by arrows. c) the Clausius-Mossotti Factor changes ($K_{CM}$) as a function of orientation for a non-spherical particle, whereas it remains constant for a sphere. Comparisons of the $K_{CM}$ values are shown below the figures for a particle with a permittivity much smaller than that of the medium (e.g. microplastic in water). d) Micrographs of ellipsoidal and spherical microplastic particle used. Scale bar: 100 μm.

**Integrated Microwave Cytometry.** The microplastic detection platform encompasses a microfluidic channel to transport the analytes, and a set of electrodes to form the microwave cytometry which possess both microwave and low-frequency sensing modalities. The electronic waveforms obtained by these modalities are processed with Machine Learning (ML) approaches to obtain the shape features of ellipsoid microparticles. To provide labelled training data for ML, a microscope is also used to observe the sensing region.

Microwave cytometry uses a low-frequency (350 kHz) resistive sensor and a microwave-band resonant capacitive sensor. The low frequency resistive sensor, commonly referred to as Coulter counter, is used to obtain the geometric diameter of dielectric particles since its operating principle is based on detecting the blockage amount in the ionic current conducted between the electrodes: the blockage in the current is proportional to the geometric volume of the particle.

The high frequency capacitive sensor is based on a split-ring microwave resonator which is used to measure the electrical size of the analytes. The split ring resonator contains two concentric rings; the outer ring receives the microwave signal from the outside electronics and excites the inner ring inductively. Due to the standing wave mode in the split region of the inner ring, high intensity electric field is generated. When a particle passes through the split region, the capacitance changes between the electrodes which causes simultaneous shifts in the amplitude and phase of the resonator.

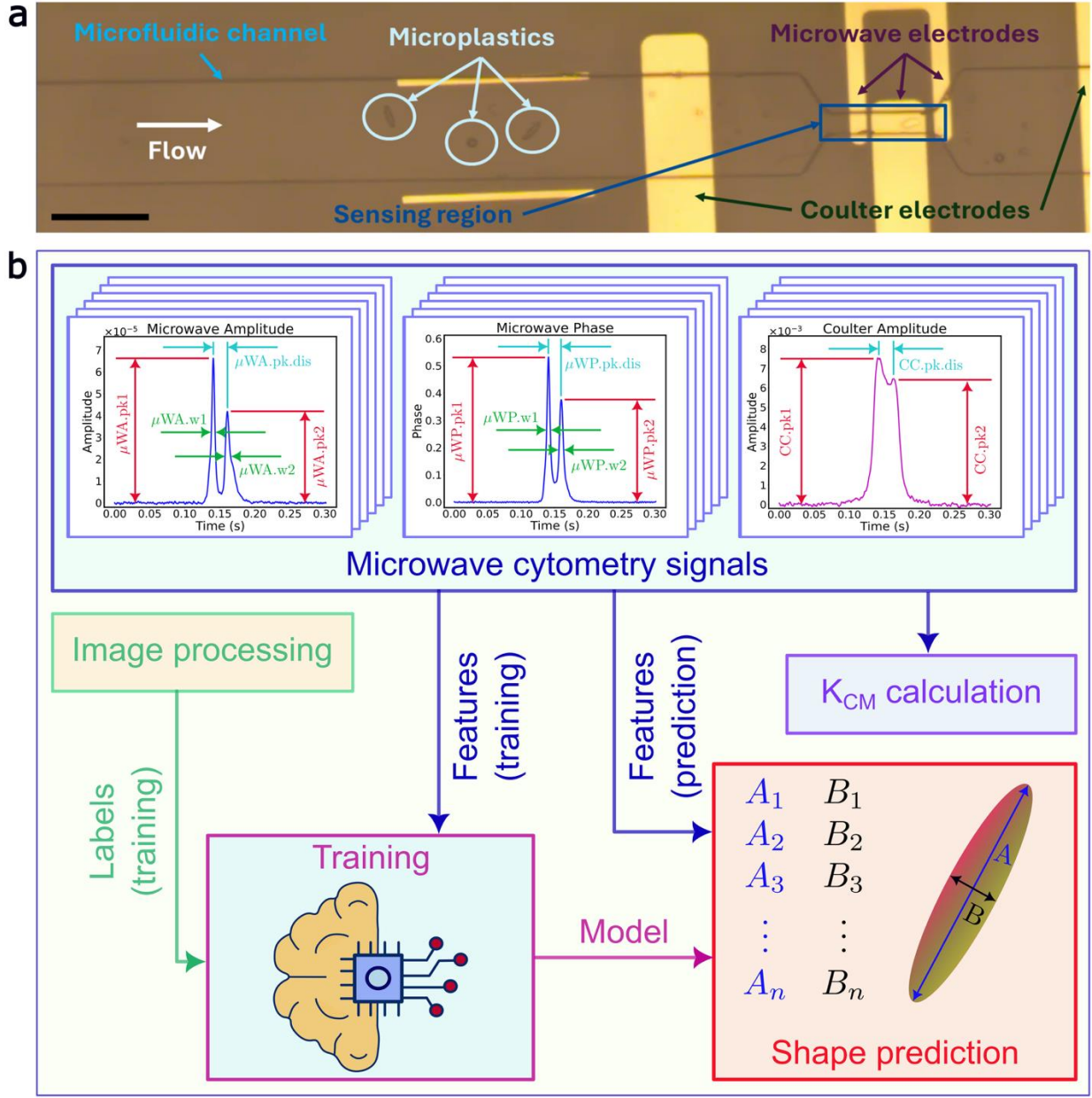

**Figure 2.** a) Micrograph of microwave cytometry device for obtaining electronic fingerprint of single particles. As a particle flow through the sensing region, it generates signal due to capacitance change and resistance changes in high frequency, and low frequency sensors, respectively. Scale bar: 200 µm. b) Methodology utilized for detecting the ellipsoidal microplastics and predicting their shape. Representative signal (after signal processing) for individual microplastics from high frequency sensor (microwave amplitude and phase) and low frequency sensor (Coulter amplitude) along with the shape descriptive features from these waveforms are shown.

The capacitance change of the resonator can be calculated by:

$$\Delta Y = \Delta R \sin(\theta) + R\cos(\theta)\Delta\theta$$

where $\Delta Y$ is the change in the out-of-phase component of the reflected voltage ($Y \equiv R \sin(\theta)$), R is the baseline amplitude, and $\theta$ is the baseline phase. We note that [31]:

$$\Delta Y \propto (\Delta C)/C$$

where C is the total capacitance of the resonator, and $\Delta C$ is the change of capacitance due to the passage of a particle. The change of phase ($\Delta\theta$) and amplitude ($\Delta R$) are obtained as individual particles pass the sensing region.

The capacitance change $\Delta C$ induced by a particle depends on its geometric volume and Clausius-Mossotti factor ($K_{CM}$). The Clausius-Mossotti factor is a function of the microwave permittivity of the particle passing and that of the liquid, along with the shape of the particle. For spherical particles, which are almost always used for microfluidic sensing, the $K_{CM}$ is given as:

$$K_{CM,SPHERE} = \frac{\varepsilon_p - \varepsilon_m}{\varepsilon_p + 2\varepsilon_m}$$

On the other hand, for an ellipsoid particle, the electrical polarizability depends on the orientation (Fig. 1). Along the axis i, the $K_{CM,i}$ is represented as [33]:

$$K_{CM,i} = \frac{1}{3}\frac{\varepsilon_p - \varepsilon_m}{\varepsilon_m + n_i(\varepsilon_p - \varepsilon_m)}$$

where $\varepsilon_p$ is the microwave permittivity of the particle, $\varepsilon_m$ is the microwave permittivity of the medium, and $n_i$ is the component of the depolarization factor along the axes of the ellipsoids (i = x, y, z). When the major axis of a prolate ellipsoid is parallel to the electric field direction, the depolarization factor is represented as

$$n_\parallel = n_x = \frac{1-e^2}{2e^3}\left[\log\left(\frac{1+e}{1-e}\right) - 2e\right]$$

Where e denotes the eccentricity, which is defined as:

$$e = \sqrt{1 - \left(\frac{B}{A}\right)^2}$$

where A and B are the major axis length and minor axis length, respectively. Due to the symmetry of the ellipsoids about the major axis, the depolarization factors in the direction of the two equal minor axes are identical and is given by

$$n_y = n_z = (1 - n_x)/2$$

since

$$n_x + n_y + n_z = 1$$

For the perpendicular orientation of the prolate ellipsoid i.e., the major axis of the prolate ellipsoid is perpendicular to the electric field direction, the depolarization factor, $n_\perp$ is same as $n_y$ and $n_z$.

Hence, the effective depolarization factor can be written as

$$n(\theta) = n_\parallel \cos^2(\theta) + n_\perp \sin^2(\theta)$$

For the simplified case of a sphere where the contribution of the depolarization factors in all directions is same i.e., $n_i = 1/3$; with this substitution, the $K_{CM}$ for sphere can be extracted:

$$K_{CM} = \frac{\varepsilon_p - \varepsilon_m}{\varepsilon_p + 2\varepsilon_m}$$

As the equations indicate, the higher the contrast between the particle and the medium, higher the capacitance change. Considering the permittivity of microplastics to be in the range of 2-4, and that of water to be around 80, a high capacitance change, therefore, a very high signal-to-noise ratio can be obtained in integrated microwave cytometry.

**Sensor Fabrication.** The sensor device fabrication includes the fabrication of the electronic sensor containing the high frequency microwave resonator and low frequency Coulter counter, and the PDMS microfluidic channel. For the electronic sensor, fused silica wafer (Quartz Unlimited LLC) of 500 μm thickness was used as the substrate. Photolithography was used to pattern the split ring resonator and the Coulter electrodes on the substrate which was then metallized with 10 nm Cr (as adhesion layer) and 150 nm Au using thermal evaporation. The final patterned coplanar electrodes were obtained after lift-off in acetone. The outer ring near the edge of the fused silica wafer was patterned to be split such that the split lines can be soldered to a subminiature A (SMA) connector to interface the microwave circuitry. The Coulter electrodes on the sensing region were extended over the resonator rings with wire-bonding or soldering for electrical connection to the low frequency circuitry. Using standard soft lithography, microfluidic channel was fabricated with polydimethylsiloxane (PDMS, Dow Chemicals) having inlet and outlet regions. The channel consisted of a constriction for the sensing region (40 μm x 45 μm x 150 μm), apart from the constriction the width of the channel was 250 μm. Finally, the microfluidic channel was aligned and bonded ($O_2$ plasma treatment) such that the constriction aligns with the sensing region of the electronic sensor (Fig. 2a). The details of the sensor design and fabrication can be found in earlier publications [31, 32, 34, 35].

**Electronic Measurement.** The shape measurement workflow (Fig. 2b) includes electronic measurement in micro-wave cytometry, automated shape measurement using image processing, and machine learning to predict the shape of individual microplastics based on electronic fingerprinting of microplastics in integrated microwave cytometry.

For the low frequency resistive sensor, a 1 V peak to peak signal at 350 kHz was used to drive one Coulter electrode using a lock-in amplifier (Zurich Instruments, HF2LI). The resulted ionic current is collected at the other electrode and converted to voltage using a trans-impedance amplifier (Zurich Instruments, HF2TA) followed by a lock-in amplifier (Zurich Instruments, HF2LI). The lock-in time constant, and the sampling rate were set as 501 µs, and 14.39 kSa/s, respectively. From the experiments, the typical duration of the particles in the sensing region were observed to be 160 ms.

For the high frequency capacitive sensor, the resonance characteristics of the split-ring resonator is first observed using a vector network analyser. Then the device is connected to a custom measurement circuit [31, 36-39] which drives the split-ring resonator at its resonance frequency (~5 GHz). In this circuit, single side band modulation was employed using two lock-in amplifiers (Zurich Instruments, MFLI) as part of the circuitry. As the operating frequency of the lock-in amplifier was lower than the resonance frequency, a custom-build single side band heterodyne circuitry was utilized this way up-conversion and down-conversion were used for reading the microwave signal. The details of the circuitry are discussed in earlier publications [31, 32].

Ellipsoidal microparticles were fabricated from spherical polystyrene (PS) particles (Polysciences) with an initial diameter of 24.9 ± 0.77 µm using a stretching procedure [40, 41]. In brief, the particles were first suspended in a polyvinyl alcohol (PVA) solution and dried to form a film. The film was then stretched approximately by a factor of 2 in an oil bath at 120 °C, causing the embedded particles to elongate. Afterwards, the film was dissolved in water and the PVA was removed through successive washing steps. The microparticles are suspended with a concentration of ~3.5 x $10^4$/ml in 0.2% Tween® 20 (Sigma-Aldrich, P1379-500ML) added Dulbecco's Phosphate Buffered Saline (biowest, L0615-500). They are transported to the sensing region using Fluigent (MFCS-EZ) pressure control system, the pressure was maintained to limit the passage of multiple particles simultaneously. The electronic measurements are conducted concurrently with the image acquisition by a microscope (Carl Zeiss) at 5x magnification (field of view 2500 µm x 2000 µm). The video from the microscope was acquired with OBS software [42] at 60 fps. The microscopic image of the sensing region of the setup, together with several ellipsoid microplastic particles, can be seen in Fig. 2a.

**Signal Processing and Feature Extraction.** Custom-written MATLAB codes were used to process features for individual microplastics from time-series microwave and Coulter signals: for each particle, the shifts in microwave amplitude, microwave phase, and Coulter amplitude were extracted. Upon resampling to 10 kSa/s, and synchronization of microwave and Coulter signals, baseline reduced signals were obtained. A built-in MATLAB function (findpeaks) was utilized to identify the individual passage of the microplastics from the microwave signals. Further, the microwave amplitude, microwave phase, and Coulter amplitudes for individual microplastics are stored after checking the valid events (e.g., events where two particles passing simultaneously from the channel were discarded as explained in detail in [32].

In this sensor design, the inner ring of the resonator consists of two gaps with unequal width (15 µm and 25 µm), thus the resultant signals have two peaks (Fig. 2b). The distance between the peaks can be used to obtain the velocity of the particle passage. Since the narrower gap will have a higher electrical field intensity, the capacitance change will also be higher, consequently a higher peak is observed compared to the second peak which is due to passage of the particle in the wider gap [43]. Coulter amplitude signals also exhibited similar two-peaked events due to the presence of microwave gold electrodes in the constriction [31].

As three different waveforms (microwave amplitude, microwave phase, Coulter amplitude) were collected for each microplastic particles (Fig. 2b), these waveforms were processed to extract indicative features. A total of 20 features were extracted and stored: 6 peak heights, 3 peak-to-peak ratios, 3 peak-to-peak distances, 4 FWHM (full width at half maximum) for microwave amplitude and phase, 2 width ratios (microwave amplitude and phase), and two intra-signal peak ratios (microwave amplitude to Coulter amplitude). FWHM could not reliably be extracted for Coulter amplitude signals due to peak overlap in a significant portion of the events.

**Image Processing for Training Labels.** Image processing was used in the training phase for generating the labels used in supervised learning of electronic signals. For image processing workflow (Fig. 3), Python3 was utilized due to the availability of libraries and functions related to image data management, image processing, and computer vision (OpenCV [44], NumPy [45], Pandas [46], Scikit-image [47], Matplotlib [48], Trackpy [49]). We employed a parallelized video processing workflow to sequentially extract frames from the recorded microscopy videos. The extracted frames were then spatially cropped to limit the region of interest to the specific region of the microfluidic channel, followed by grayscale conversion for reducing computational complexities while retaining structural information for downstream image processing.

Background subtraction technique was used to isolate the moving microplastics in the microfluidic channel from the static background (i.e., PDMS wall, gold electrode). In this step, adaptive Gaussian mixture model-based pixel-level background subtraction [50] was utilized to account for uneven illumination due to the presence of gold electrodes in the region of interest. The background subtraction was applied frame-by-frame to generate foreground images containing only the moving microplastics.

To denoise the foreground images and to preserve the structural features e.g., edge of the microplastics in them, total variation denoising was used which minimizes the total variation of an image [51]. Canny edge detection algorithm [52] was then applied on the denoised images to delineate the boundaries of the microplastics.

Morphological filtering operation, a standard image processing technique, was applied to remove any possible noise-induced small, interior gaps within the image structure. Then, contours are detected and filtered based on minimum and maximum areas to avoid any potential artefacts, and the filtered contours are overlayed on a blank greyscale mask, this mask now contains the labelled segmented microplastics. For measurement of the shape characteristics, these segmented microplastics are approximated as ellipsoids since heterogenous microplastics from environmental samples can reasonably be represented as such geometry [53]. Geometric parameters (e.g., area, centroid, major and minor axis lengths) are extracted from individually connected components i.e., individual microplastics, with appropriate scaling factors. The major and minor axis lengths are then visualized on the particles in the original cropped frames with text annotations.

As microplastics flow in the microfluidic channel, the same particles are observed in a significant set of consecutive frames and their shape characteristics measured. It is crucial to identify and link the same particles along with their shape characteristics across frames in time-lapse image sequences. Trackpy [49], an implementation of Crocker, and Grier [54] algorithm, were used to detect the particles in each frame, to associate the detected particles across frames along with their measurements, and to provide a unique ID based on spatial proximity and register the exit of the particle. The ensemble of the shape characteristics measurement for individual microplastics are stored.

As the uneven illumination on the Coulter electrode region can result in distortion of the particles in foreground images, the shape characteristics measured for particles on top of the left Coulter electrode are discarded from further processing. When considering the measurements of major and minor axis lengths of a single microplastic, outlier values that are too high or too low compared to the rest can skew the average. To address this, we considered trimmed mean as a measure of central tendency thereby we have removed the lowest 10% and highest 10% measurements (for the same particle in different frames) from the ensemble and then calculated the ensemble average to be the true major and minor axis lengths of the microplastic. Finally, the electronic signals and measured shapes are cross matched to ensure the individual microplastics correspond to the correct electronic measurements and shapes for the ma-chine learning workflow.

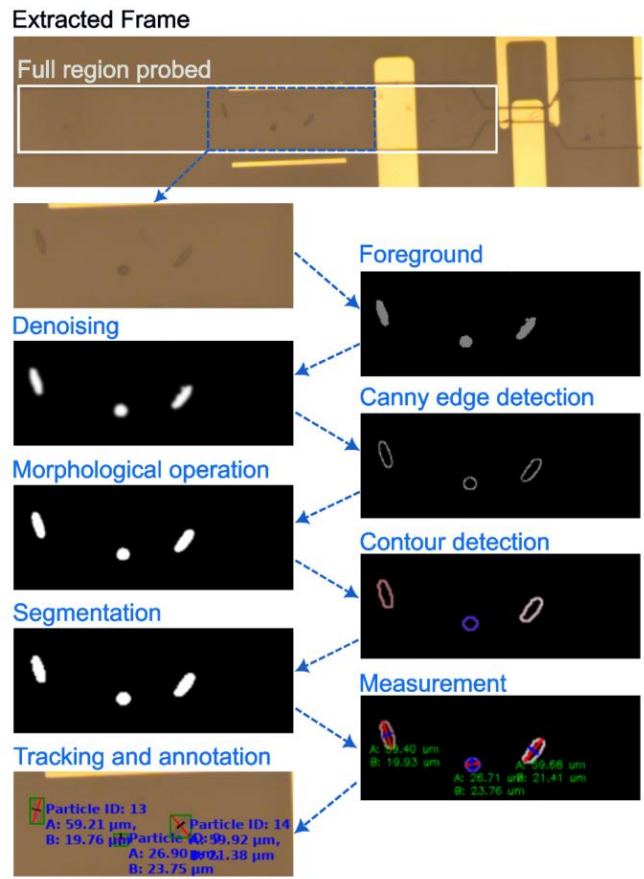

**Figure 3.** In the training phase, ground truth was obtained through an image processing workflow to obtain the major and minor axis lengths, of each particle as they flow through the microfluidic channel. The labels obtained in this step are later used for training the machine learning workflow. White rectangle in 'extracted frame' indicates the full region probed with this method, blue dashed region shows the zoom-in view region in latter steps.

**Machine Learning Model.** Supervised machine learning provides a suitable framework for predicting the shape of the microplastics due to its ability to model complex, non-linear relationships between multidimensional input features extracted from the electronic measurements in integrated microwave cytometry and labelled targets. For developing the machine learning workflow, we have utilized the machine learning libraries in Python3 e.g., SciPy [55], Scikit-learn [56], NumPy [45], Pandas [46], that enabled efficient data handling, streamlined machine learning algorithms.

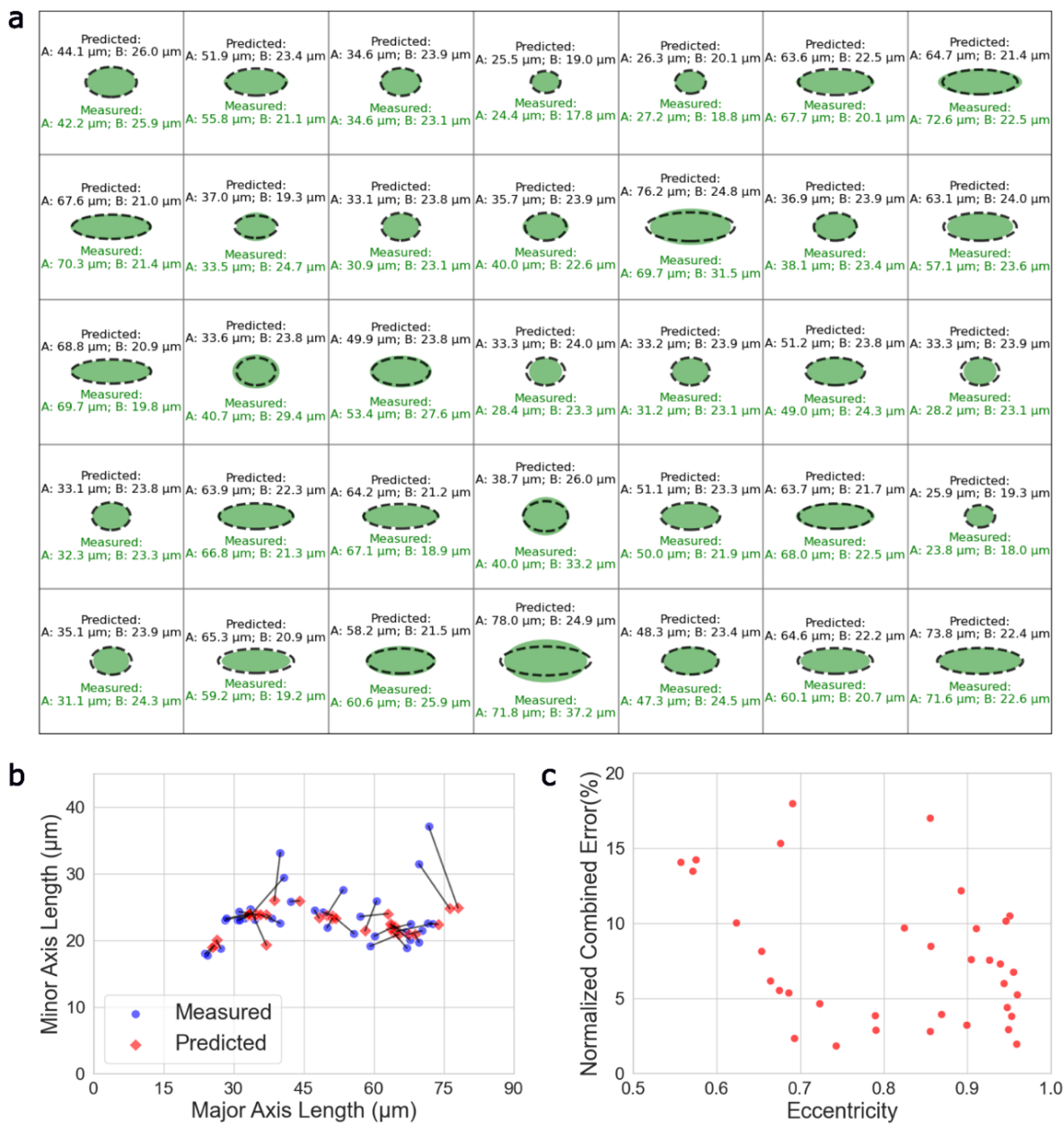

**Figure 4.** a) Ellipses overlayed for the test data set using optically measured (green, filled ellipses) and electronically predicted (ellipses with black, dashed outlines) axis lengths. b) Comparison of measured and predicted major and minor axis lengths for test data set, with each pair connected to illustrate prediction deviation, c) normalized combined error is shown as a function of eccentricity, the mean normalized combined error was calculated as 7.6%.

The dataset for machine learning workflow contained the features extracted from electronic measurements (input features) of 175 individual microplastics (eccentricity range: 0.55-0.99, mean: 0.8), and their shape measurements (target labels) obtained from image processing. Both features and labels were normalized (z-score normalization $z=(x-\mu)/\sigma$) to ensure zero mean and unit variance such that the scaling heterogeneity among the features can be eliminated [57], this normalization also assists in numerical stability for machine learning algorithms. Following normalization, the data was split into training and testing set (80%-20%), a standard process in machine learning workflow, with a fixed random state to ensure reproducibility. This way 80% of the data was used for training the algorithm, and the remining 20% of the data was isolated to avoid data leakage and was used later to assess the predictive accuracy of the model. Data leakage can take place when the test data is also included in the training data such that the algorithm memorizes instead of generalizing the model on the patterns.

A random forest regression model was utilized to capture the relationship between the input features and target labels by aggregating the outputs of multiple decision trees [58]. Random forest regression models are suitable due to their ability to handle high-dimensional feature space, and to capture complex, non-linear interactions among features [59].

For robust model generalization, a k-fold cross-validation was employed by incorporating random shuffling for optimizing model generalization. This process divides the dataset into k subsets, trains the model iteratively on k-1 folds while testing on the remaining fold, to assess the model performance, and the generalization of the model across different data splits.

A grid search approach was used for hyperparameter optimization over a parameter space. Hyperparameters, set prior to training, are referred to as the configurable variables in ML algorithms that govern the model architecture, complexity, and training dynamics. For a random forest regression model, typical hyperparameters are number of trees, tree depth, minimum number of samples at a leaf node, and minimum number of samples to split. Mean absolute percentage error was used as the scoring metric for the optimization to enhance predictive accuracy. This step enabled tuning the hyperparameters that generalize well across different validation folds.

The optimal hyperparameters were used to refine the random forest regression model wrapped within a multi-output regression framework for major and minor axis length prediction. Upon evaluation of the optimized model on the folds, the model was finally used for training on the full training data set and evaluated on the test data set to assess its generalization on the unseen data. The predictions by the model were then inverse transformed to their original scales and absolute percentage errors for individual particles were calculated. The mean normalized combined error was calculated as:

$$Normalized\ combined\ error = \frac{\sqrt{(A_{\text{pred}} - A_{\text{true}})^2 + (B_{\text{pred}} - B_{\text{true}})^2}}{\sqrt{A_{\text{true}}^2 + B_{\text{true}}^2}}$$

Where A denotes the major axis length, and B the minor axis length. The subscript pred stands for predicted and true for the ground truth obtained by the microscope image. The overall normalized combined error was found to be 7.6 % for prediction on the test set. Fig. 4 shows the visual representation of the prediction of the shape of the ellipsoidal microplastics and their deviation from the true values as obtained from the image processing (Fig. 4 a, b), these represent the microplastics from the test set that was unknown to the ML algorithm. It can be observed that the machine learning model can predict the major and minor axis lengths of ellipsoidal microplastics with reasonable accuracy for all eccentricity values. No distinct systematic correlations between the microplastics shape (based on eccentricity) and the accuracy of predicted major and minor axis lengths were observed (Fig. 4c), indicating the model's performance was not limited by the eccentricity of the particles. The accuracy in predicting the shape of the ellipsoidal microplastics shows the electronic signals as obtained in the integrated microwave cytometry to indeed contain the shape descriptors of the individual microplastics. The prediction accuracy of the developed machine learning model can be affected by a variety of factors which are discussed in a subsequent section.

**Effect of Eccentricity and Orientation on Clausius-Mossotti Factor.** In addition to the measurement of ellipsoidal microplastics, electronic measurement was also carried out for 25 μm spherical particles (n = 171) as independent reference experiments. The $K_{CM}$ values were calculated from the microwave and Coulter signals for both ellipsoidal and spherical microplastics [31]. The ellipsoidal microplastics exhibit a wider distribution in $K_{CM}$ values compared to spherical microplastics, moreover, the mean $K_{CM}$ value of the ellipsoidal microplastics shifted from the mean $K_{CM}$ value of spherical microplastics by 24.1% (Fig. 5c). The coefficient of variation for the experimentally obtained $K_{CM}$ values were found to be 4.9% and 22.6% for the spherical and ellipsoidal microplastics, respectively. The relatively larger dispersion for the ellipsoidal particles can be attributed to its dependency on the polarization factor, $n_i$. The polarization factor $n_i$ depends on the eccentricity and the direction e.g., the angle between the major axis and the electrical field direction. As a result, an ellipsoid can have various orientations while passing through the sensing region with respect to the electric field present, hence can exhibit variations in polarizability due to the orientation alone —in contrast to spherical particles. Moreover, the polarizability will depend on the different degrees of eccentricity for the ellipsoidal microplastics in the sample, which at this stage have inevitable dispersion.

To understand the dispersity better, the variation was also studied through Monte Carlo simulations to obtain $K_{CM}$ values by having both random eccentricity and orientation for polystyrene ellipsoidal particles. The generation of random eccentricity values for Monte Carlo simulation was done using histogram resampling such that the distribution of the eccentricity of ellipsoidal microplastics used in experiments (Fig. S4) is maintained. Constraints were applied in the simulation to mimic the experimental conditions e.g., restricting the proximities of the vertical and horizontal orientation of the ellipsoids due to channel height and width constraints. In reflection with the experimental case, the mean $K_{CM}$ values of ellipsoidal microplastics (e: 0.55-0.99) are shifted downwards by 11% from that of spherical microplastics (Fig. 5d). In the simulation, the coefficient of variation for $K_{CM}$ in ellipsoidal microplastics was observed to be 7.6%. Simulation results recapitulate the fact that the deviation in $K_{CM}$ from that of a sphere increases as eccentricity increases (Fig. S3). The observation from the Monte Carlo simulation further validates the experimental observations on the strong effect of eccentricity and orientation on the Clausius-Mossotti factor.

While the analytical $K_{CM}$ bounds (Fig. 5b) progressively diverge with eccentricity, it should be noted that, this calculation in Monte Carlo simulation assumes uniform prolate particles, whereas experiments also included curved prolate particles having non-uniformity along major axis (e.g., pear or banana shaped) which would possibly elucidate the increase in the coefficient of variation, along with other experimental uncertainties. Interestingly, when average $K_{CM}$ along the bound for eccentricity variation is considered i.e., average $K_{CM}$ values include all the possible orientations for that eccentricity, it shows slight shift from the $K_{CM}$ of a sphere (Fig. S2). The analytical $K_{CM}$ values for the prolate ellipsoids coincide with that of sphere when the major axis is at an angle of $54.7^0$, so called 'magic angle' [60] with the electric field direction (Fig. 5b, Fig. S2) as the anisotropic contributions of the ellipsoids equal to the isotropic response of a sphere. While the average $K_{CM}$ line shifts from that of sphere gradually, in the eccentricity range of 0.90-0.99, the average $K_{CM}$ line rises sharply. The results from Monte Carlo simulations, when disregarding the constraints imposed, also lie in close proximity to theses average $K_{CM}$ values for the same reason i.e., orientation averaged values. However, in experimental conditions and in the absence of any other force, the orientation of the particles will depend on the hydrodynamic forces acting on it. Hence, the $K_{CM}$ values of the prolate ellipsoids are expected to show greater shift from the average $K_{CM}$ values since the particles of the same shape will not experience all the orientations due to applied hydrodynamic forces as the orientation and trajectory is highly sensitive to the shape of the particle and their interactions with the flow field, especially for axisymmetric particles [61].

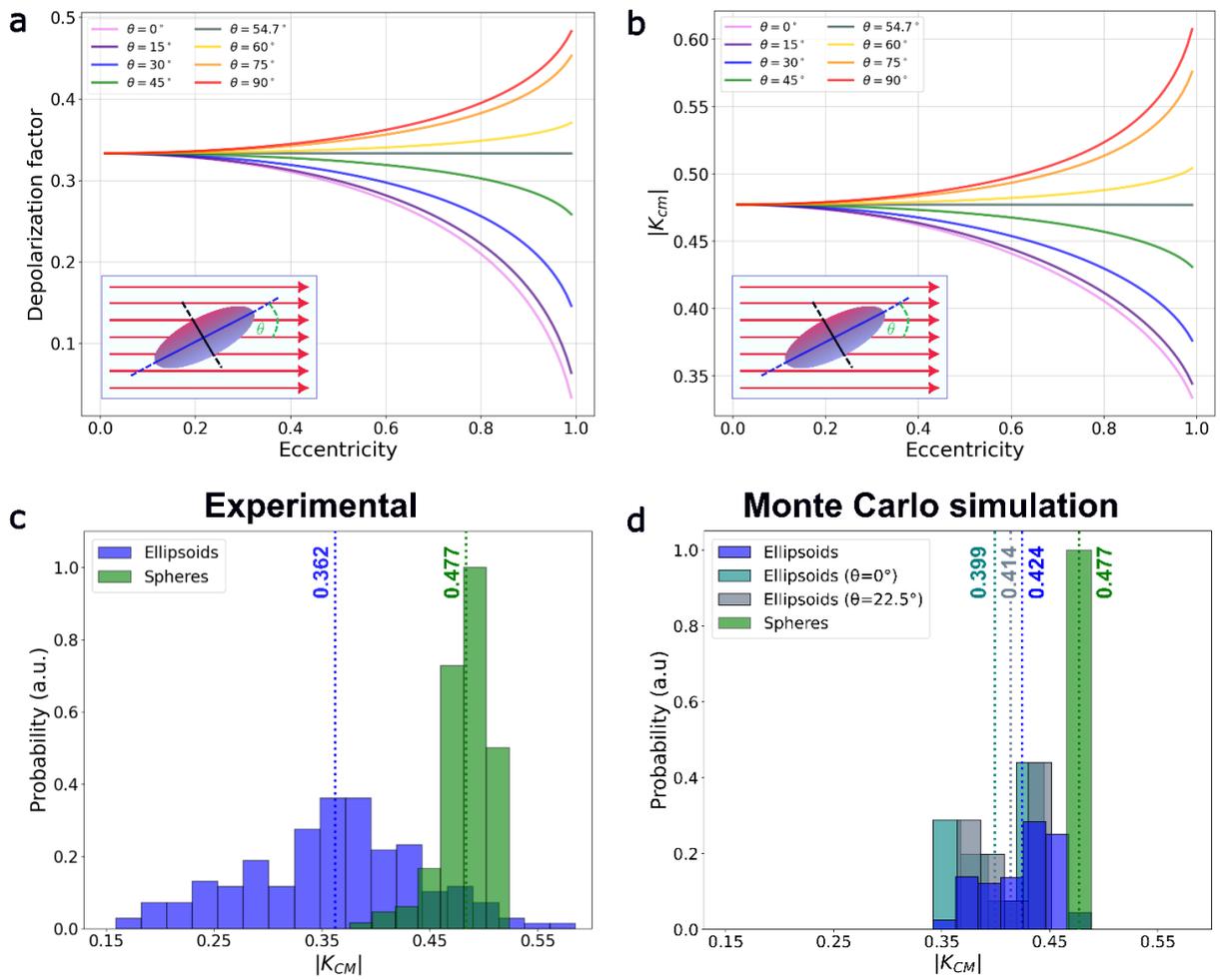

**Figure 5.** Analytical bounds for (a) depolarization factor and (b) $|K_{CM}|$ show divergence with increase in eccentricity. For a sphere (e = 0), the depolarization factor is 0.333, the bound diverges with eccentricity increase and reaches to 0.483 (for perpendicular orientation) and 0.034 (for parallel orientation) at e = 0.99. For a sphere, the $K_{CM}$ value is 0.477. At e = 0.99, the $|K_{CM}|$ values at the bounds are 0.583 and 0.350 for the perpendicular and parallel orientation of the ellipsoid, respectively. For $K_{CM}$ calculation, $\varepsilon_p$ = 2.4 (Polystyrene) and $\varepsilon_m$ = 78 (water) were considered. Histogram of Clausius Mossotti factor obtained from (c) experiments and (d) Monte Carlo simulation show wider and shifted $K_{CM}$ distribution for ellipsoidal (e > 0.55) polystyrene microplastics at random orientations compared to their spherical counterparts. At fixed orientations ($\theta = 0^0$ and $\theta = 22.5^0$) potentially imposed by the fluid flow field, $K_{CM}$ distribution of the ellipsoids show further deviations from that of sphere. The dotted lines represent the mean values. To facilitate comparison in the histogram, the mean experimental $|K_{CM}|$ value for spheres (measured in a.u.) was scaled to 0.477, the analytical $|K_{CM}|$ value for polystyrene spheres. The same scaling was applied to all experimental $|K_{CM}|$ values for spheres and ellipsoids.

The classical theory by Jeffery suggests that the prolate ellipsoids will tend to orient themselves with the major axis parallel to the fluid flow direction ($\theta \approx 0^0$) in a uniform shear flow field by tracing the Jeffery orbits; depending on the orbital constant the prolate ellipsoids will exhibit tumbling, kayaking, log-rolling, or cartwheeling motions [62-64]; in confined flow, they tend to oscillate[65] around $\theta \approx 0^0$. Experimental observations report a deviation from the theoretical preferential orientation to slight inclination ($\theta \approx 22.5^0$) which could potentially be due to the wall effect, initial orientation, interparticle interaction, non-uniform shear, and inertial corrections [64]. Prolate ellipsoids can exhibit other orientations with complex tumbling as well [66]. Hence, the prolate ellipsoidal microplastics in the microchannel will tend to exhibit orientation with its major axis parallel or with slight inclination to the flow field in the channel. When the particles enter the constriction, their dynamics will change due to increased local shear and wall effect, however, depending on the shape this effect will further align the particles with the flow direction due to tighter confinement, especially for particles with higher eccentricity values [65], and similarly to the electric field direction. Therefore, the $K_{CM}$ values are expected to be closer to the lower bound (Fig. 4b, Fig. S2), i.e., parallel orientation of the prolate ellipsoids with respect to the electric field and will show further shift from the $K_{CM}$ value of the spheres if the orientation bias with the hydrodynamic field interactions considered. Monte Carlo simulation results indicate that when the orientation is restricted to $\theta = 0^0$ and $\theta = 22.5^0$ to imitate the theoretical consideration [65] and the experimental observation [64] of the prolate ellipsoid motion, the resulting $K_{CM}$ shows deviation of 16.35%, and 13.21%, respectively, from that of the sphere (Fig. 5d). It is to be noted that, however, these conclusions are valid for axisymmetric prolate particles. The presence of non-axisymmetric particle further complicates these dynamics and can thus deviate more than the axisymmetric cases.

**Discussion: Factors Affecting Prediction Accuracy.** There could be various factors that can affect the prediction accuracy of the shape measurement of the ellipsoidal microplastics. They can be originated from the uncontrolled rotation of the particles in the sensing region, limitations in electronic sensing performance, inaccurate labelling of particles during the use of image processing for training the machine learning model, or from the machine learning model itself.

Due to the presence of coplanar electrodes in the sensing region, non-uniform electrical fields exist for both high frequency and low frequency sensors, resulting in vertical position dependency of the signal. Hence, for the same shaped microplastic passing through different heights of the microfluidic channel in the sensing region, there will be variations in the signal amplitudes. This positional dependency potentially added uncertainty to some of the features extracted from the sensor signals. Also, the high-speed tumbling, rotational drifting or lateral displacement of the microplastics while passing through the sensing region can also alter the signal to some extent.

The coefficient of variation for the $K_{CM}$ of the sphere is taken to indicate the effect of height variations in the channel, the experimental results show a coefficient of variation of 4.9% for the $K_{CM}$ for the monodispersed spherical particles. When ellipsoids (e > 0.55) are considered, the coefficient of variation for $K_{CM}$ was observed to be 22.6% indicating strong effects of shape and orientation compared to spheres while also incorporating the height dependency. From the analytical model-based Monte Carlo simulations having experimentally observed constraints, we observed coefficient of variation of the $K_{CM}$ reaches to 7.6% as compared to 22.6% for experimental observations. It can be hypothesized that the high-speed, complex motion in the sensing region due to the shape of non-spherical particles affects the accuracy markedly in compared to the height difference alone. Indeed, prolate ellipsoids exhibit complex translational and rotational dynamics in low Reynolds number pressure driven flow [64]. However, the notable difference of coefficient of variation can be attributed to other factors along with the experimental uncertainty. The analytical model considers perfectly symmetrical prolate ellipsoids, as noted before, minority of particles tested showed limited yet noticeable asymmetry. The signals from these particles are not expected to be encapsulated by $K_{CM}$ bounds as deduced from the analytical expressions, hence leading to greater variations in the experimental observations.

There could be several sources of uncertainty and potential errors during automated measurement of microplastics in a microfluidic channel with image processing workflow which is later used for training the machine learning model. The fast movement of some particles can potentially lead to pixelation artefacts, resolution loss, and motion blur [67], when combined with background subtraction and denoising steps, they may distort particle shapes or lead to over/underestimation of the particle sizes. The asymmetry and complex motion of ellipsoidal microplastics (e.g., tumbling, rotation, translational movement) especially when out-of-focus, can also alter contour detection and segmentation accuracy and thus introduce additional sources of uncertainty for shape measurement. Orientational variability may introduce measurement bias in each frame since the 2D projected major and minor axis lengths depend on the instantaneous orientation relative to the imaging plane [22]; due to tilt or rotation the projected major axis may appear shorter while the minor axis may appear artificially elongated. Since many of the particles are asymmetric, if any microplastic does not exhibit all orientations during observation, the ensemble average may fail to represent shape extremes. Relative smaller size of the particles compared to the field of view, optical aberrations, limited contrast and non-uniform illumination due to presence of reflective gold electrodes can further hinder precise particle edge delineation.

Random forest regression, although a robust supervised machine learning algorithm, can sometimes have relatively lower extrapolation accuracy especially when the trees have fewer examples to learn from [68]. Hence the model may predict shape with higher errors when the particle's motion or shape is not well-represented in the range of trained target values. Since the model tends to minimize error by focusing on regions having large number of samples, it may cause underfitting for the sparsely sampled region of target variables.

Accounting for the error sources which may cause marked deviation in the estimation of the particle shape at micro scale, the resulting predicted particle shape shows reasonable accuracy with a mean normalized combined error of 7.6%. Additionally, the normalized combined error (Fig. 4c) does not show any systematic correlations with eccentricity. This indicates a good prediction capability of the machine learning model to predict the particle shapes based on the electronic fingerprinting of individual microplastics in integrated microwave cytometry.

**Conclusion**

In this work, we detected non-spherical polystyrene mi-croplastic particles using integrated microwave cytometry and calculated Clausius-Mossotti factor for individual microplastics. We develop an image processing workflow to automate the training phase, i.e., label extraction measurements of these microplastics as they flow through the microfluidic channel. This way we obtained the shape parameters as labels used to train the machine learning algorithm. Using features from the electronic signal waveform as obtained from the integrated microwave cytometry, we predict the shape of individual microplastics with high accuracy utilizing machine learning model. This study, to the best of our knowledge, is the first report of all-electronic single-particle-level detection and shape measurement of non-spherical microplastics in a flow-through manner.

To precisely obtain the material permittivity, it is essential to know the eccentricity and the orientation of the particle with respect to the electric field. While eccentricity can be obtained from the shape measurement using the developed machine learning algorithm, obtaining the exact orientation of the particle with respect to the electric field is uncertain due to the rotation of the particles in the sensing region, as well as the presence of the coplanar electrode resulting a non-uniform non-directional electric field. A size and shape invariant focusing of particles to a sensing region having a unidirectional and uniform electric field e.g., as obtained using 3D electrodes [32, 35], can have the capability to obtain highly accurate material permittivity values of individual microplastics with integrated microwave cytometry. This can potentially close the gap between the requirements of environmental science and the current effort by researchers to provide lab-on-a-chip solutions for environmentally relevant non-uniform non-spherical microplastics detection and their shape measurement.

## AUTHOR INFORMATION

**Corresponding Author**

M. Selim Hanay: selimhanay@bilkent.edu.tr

**Author Contributions**

Conceptualization: SS, UT, and MSH; methodology: SS, SSUA, UT, LW and MSH; software: SS, and SSUA; validation: MSH; formal analysis: SS, and SSUA; investigation: SS, and SSUA; resources: LW, and MSH; data curation: SS; writing, original draft preparation: SS, and SSUA; writing, review and editing: LW, and MSH; visualization: SS, and SSUA; supervision: MSH; project administration: MSH; funding acquisition: MSH. All authors have read and agreed to the published version of the manuscript.

**Funding Sources**

This project has received funding from the European Research Council (ERC) under the European Union's Horizon Europe program by Proof-of-Concept Grant RAMP-UP, Grant No: 101113438. MSH acknowledged support from FCT through ERC-PT Careers program.

**Acknowledgements**

The authors thank Maruf A Dhali for useful discussions related to image processing.

**Notes**

All data and code are uploaded to Zenodo repository.

## REFERENCES

(1) Rochman, C. M., Microplastics research—from sink to source. *Science* **2018,** *360* 6384, 28-29.
(2) Thompson, R. C.; Courtene-Jones, W.; Boucher, J.; Pahl, S.; Raubenheimer, K.; Koelmans, A. A., Twenty years of microplastic pollution research—what have we learned? *Science* **2024,** *386* 6720, eadl2746.
(3) Vethaak, A. D.; Legler, J., Microplastics and human health. *Science* **2021,** *371* 6530, 672-674.
(4) Cincinelli, A.; Scopetani, C.; Chelazzi, D.; Lombardini, E.; Martellini, T.; Katsoyiannis, A.; Fossi, M. C.; Corsolini, S., Microplastic in the surface waters of the Ross Sea (Antarctica): occurrence, distribution and characterization by FTIR. *Chemosphere* **2017,** *175*, 391-400.
(5) Käppler, A.; Fischer, D.; Oberbeckmann, S.; Schernewski, G.; Labrenz, M.; Eichhorn, K.-J.; Voit, B., Analysis of environmental microplastics by vibrational microspectroscopy: FTIR, Raman or both? *Analytical and bioanalytical chemistry* **2016,** *408* 29, 8377-8391.
(6) De Frond, H.; Hampton, L. T.; Kotar, S.; Gesulga, K.; Matuch, C.; Lao, W.; Weisberg, S. B.; Wong, C. S.; Rochman, C. M., Monitoring microplastics in drinking water: An interlaboratory study to inform effective methods for quantifying and characterizing microplastics. *Chemosphere* **2022,** *298*, 134282.
(7) De Frond, H.; Cowger, W.; Renick, V.; Brander, S.; Primpke, S.; Sukumaran, S.; Elkhatib, D.; Barnett, S.; Navas-Moreno, M.; Rickabaugh, K., What determines accuracy of chemical identification when using microspectroscopy for the analysis of microplastics? *Chemosphere* **2023,** *313*, 137300.
(8) Nava, V.; Frezzotti, M. L.; Leoni, B., Raman spectroscopy for the analysis of microplastics in aquatic systems. *Applied Spectroscopy* **2021,** *75* 11, 1341-1357.
(9) Schymanski, D.; Goldbeck, C.; Humpf, H.-U.; Fürst, P., Analysis of microplastics in water by micro-Raman spectroscopy: Release of plastic particles from different packaging into mineral water. *Water research* **2018,** *129*, 154-162.
(10) Ourgaud, M.; Phuong, N. N.; Papillon, L.; Panagiotopoulos, C.; Galgani, F.; Schmidt, N.; Fauvelle, V.; Brach-Papa, C.; Sempéré, R., Identification and quantification of microplastics in the marine environment using the laser direct infrared (LDIR) technique. *Environmental Science & Technology* **2022,** *56* 14, 9999-10009.
(11) Bernard, N.; Metian, M.; Oberhaensli, F.; Sznaider, F.; Ruberto, L.; Vodopivez, C.; Mac Cormack, W.; Hernandez, C. A., Identification and quantification of microplastics in the Antarctic coastal waters using laser direct infrared (LDIR). *Marine Pollution Bulletin* **2025,** *221*, 118534.
(12) Du, S.; Liu, Z.; Wu, L.; Tao, F., Identification and characterization of microplastics released during the actual use of disposable cups using laser direct infrared imaging. *Analyst* **2025,** *150* 5, 989-997.

(13) Dong, H.; Wang, X.; Niu, X.; Zeng, J.; Zhou, Y.; Suona, Z.; Yuan, Y.; Chen, X., Overview of analytical methods for the determination of microplastics: Current status and trends. *TrAC Trends in Analytical Chemistry* **2023**, *167*, 117261.
(14) Lahiri, S. K.; Azimi Dijvejin, Z.; Golovin, K., Polydimethylsiloxane-coated textiles with minimized microplastic pollution. *Nature Sustainability* **2023**, *6* 5, 559-567.
(15) Ward, E.; Gordon, M.; Hanson, R.; Jantunen, L. M., Modelling the effect of shape on atmospheric microplastic transport. *Atmospheric Environment* **2024**, *326*, 120458.
(16) Zhao, S.; Kvale, K. F.; Zhu, L.; Zettler, E. R.; Egger, M.; Mincer, T. J.; Amaral-Zettler, L. A.; Lebreton, L.; Niemann, H.; Nakajima, R., The distribution of subsurface microplastics in the ocean. *Nature* **2025**, *641* 8061, 51-61.
(17) Tatsii, D.; Bucci, S.; Bhowmick, T.; Guettler, J.; Bakels, L.; Bagheri, G.; Stohl, A., Shape matters: Long-range transport of microplastic fibers in the atmosphere. *Environmental science & technology* **2023**, *58* 1, 671-682.
(18) Preston, C. A.; McKenna Neuman, C. L.; Aherne, J., Effects of shape and size on microplastic atmospheric settling velocity. *Environmental Science & Technology* **2023**, *57* 32, 11937-11947.
(19) Kim, O.-H.; Chang, E. S.; Kang, H.; Lee, H. J., Unveiling the impact of microplastics: a perspective on size, shape, and composition in human health. *Molecular & Cellular Toxicology* **2025**, 1-10.
(20) Lozano, Y. M.; Lehnert, T.; Linck, L. T.; Lehmann, A.; Rillig, M. C., Microplastic shape, polymer type, and concentration affect soil properties and plant biomass. *Frontiers in plant science* **2021**, *12*, 616645.
(21) Lehmann, A.; Leifheit, E. F.; Gerdawischke, M.; Rillig, M. C., Microplastics have shape-and polymer-dependent effects on soil aggregation and organic matter loss–an experimental and meta-analytical approach. *Microplastics and Nanoplastics* **2021**, *1* 1, 7.
(22) Kim, J.; Kim, J. Y.; Kim, Y.; Lee, S. J.; Kim, J. M., Shape measurement of ellipsoidal particles in a cross-slot microchannel utilizing viscoelastic particle focusing. *Analytical chemistry* **2017**, *89* 17, 8662-8666.
(23) Rozman, U.; Kalčíková, G., Seeking for a perfect (non-spherical) microplastic particle–the most comprehensive review on microplastic laboratory research. *J. Hazard. Mater* **2022**, *424* 127529, 10.1016.
(24) Shiratori, K.; Bishop, L. D.; Ostovar, B.; Baiyasi, R.; Cai, Y.-Y.; Rossky, P. J.; Landes, C. F.; Link, S., Machine-learned decision trees for predicting gold nanorod sizes from spectra. *The Journal of Physical Chemistry C* **2021**, *125* 35, 19353-19361.
(25) Glaubitz, C.; Bazzoni, A.; Ackermann-Hirschi, L.; Baraldi, L.; Haeffner, M.; Fortunatus, R.; Rothen-Rutishauser, B.; Balog, S.; Petri-Fink, A., Leveraging machine learning for size and shape analysis of nanoparticles: a shortcut to electron microscopy. *The Journal of Physical Chemistry C* **2023**, *128* 1, 421-427.
(26) D'Orazio, M.; Reale, R.; De Ninno, A.; Brighetti, M. A.; Mencattini, A.; Businaro, L.; Martinelli, E.; Bisegna, P.; Travaglini, A.; Caselli, F., Electro-optical classification of pollen grains via microfluidics and machine learning. *IEEE Transactions on Biomedical Engineering* **2021**, *69* 2, 921-931.
(27) Kokabi, M.; Tayyab, M.; Rather, G. M.; Pournadali Khamseh, A.; Cheng, D.; DeMauro, E. P.; Javanmard, M., Integrating optical and electrical sensing with machine learning for advanced particle characterization. *Biomedical Microdevices* **2024**, *26* 2, 25.
(28) Reale, R.; De Ninno, A.; Nepi, T.; Bisegna, P.; Caselli, F., Extensional-flow impedance cytometer for contactless and optics-free erythrocyte deformability analysis. *IEEE Transactions on Biomedical Engineering* **2022**, *70* 2, 565-572.
(29) Jarmoshti, J.; Siddique, A. B.; Rane, A.; Mirhosseini, S.; Adair, S. J.; Bauer, T. W.; Caselli, F.; Swami, N. S., Neural Network-Enabled Multiparametric Impedance Signal Templating for High throughput Single-Cell Deformability Cytometry Under Viscoelastic Extensional Flows. *Small* **2025**, *21* 5, 2407212.
(30) Caselli, F.; Reale, R.; De Ninno, A.; Spencer, D.; Morgan, H.; Bisegna, P., Deciphering impedance cytometry signals with neural networks. *Lab on a Chip* **2022**, *22* 9, 1714-1722.
(31) Tefek, U.; Sari, B.; Alhmoud, H. Z.; Hanay, M. S., Permittivity-Based Microparticle Classification by the Integration of Impedance Cytometry and Microwave Resonators. *Advanced Materials* **2023**, 2304072.
(32) Alatas, Y. C.; Tefek, U.; Salehin, S.; Alhmoud, H.; Hanay, M. S., Rapid Differentiation between Microplastic Particles Using Integrated Microwave Cytometry with 3D Electrodes. *ACS sensors* **2025**, *10* 3, 1729-1735.
(33) Castellarnau, M.; Errachid, A.; Madrid, C.; Juarez, A.; Samitier, J., Dielectrophoresis as a tool to characterize and differentiate isogenic mutants of Escherichia coli. *Biophysical journal* **2006**, *91* 10, 3937-3945.
(34) Secme, A.; Tefek, U.; Sari, B.; Pisheh, H. S.; Uslu, H. D.; Çalışkan, Ö. A.; Kucukoglu, B.; Erdogan, R. T.; Alhmoud, H.; Sahin, O., High-Resolution Dielectric Characterization of Single Cells and Microparticles Using Integrated Microfluidic Microwave Sensors. *IEEE Sensors Journal* **2023**, *23* 7, 6517-6529.
(35) Alatas, Y. C.; Tefek, U.; Sari, B.; Hanay, M. S., Microwave Resonators Enhanced With 3D Liquid-Metal Electrodes for Microparticle Sensing in Microfluidic Applications. *IEEE Journal of Microwaves* **2023**, 1-8.
(36) Ferrier, G. A.; Romanuik, S. F.; Thomson, D. J.; Bridges, G. E.; Freeman, M. R., A microwave interferometric system for simultaneous actuation and detection of single biological cells. *Lab on a Chip* **2009**, *9* 23, 3406-3412.
(37) Nikolic-Jaric, M.; Romanuik, S.; Ferrier, G.; Bridges, G.; Butler, M.; Sunley, K.; Thomson, D.; Freeman, M., Microwave frequency sensor for detection of biological cells in microfluidic channels. *Biomicrofluidics* **2009**, *3* 3, 034103.
(38) Afshar, S.; Salimi, E.; Braasch, K.; Butler, M.; Thomson, D. J.; Bridges, G. E., Multi-frequency DEP cytometer employing a microwave sensor for dielectric analysis of single cells. *IEEE Transactions on Microwave Theory and Techniques* **2016**, *64* 3, 991-998.

(39) Kelleci, M.; Aydogmus, H.; Aslanbas, L.; Erbil, S. O.; Hanay, M. S., Towards microwave imaging of cells. *Lab on a Chip* **2018,** *18* 3, 463-472.
(40) Ho, C.; Keller, A.; Odell, J.; Ottewill, R., Preparation of monodisperse ellipsoidal polystyrene particles. *Colloid and Polymer Science* **1993,** *271* 5, 469-479.
(41) Weirauch, L.; Giesler, J.; Baune, M.; Pesch, G. R.; Thöming, J., Shape-selective remobilization of microparticles in a mesh-based DEP filter at high throughput. *Separation and Purification Technology* **2022,** *300*, 121792.
(42) Kleinjan, M. S.; Buchta, W. C.; Ogelman, R.; Hwang, I.-W.; Kuwajima, M.; Hubbard, D. D.; Kareemo, D. J.; Prikhodko, O.; Olah, S. L.; Wulschner, L. E. G., Dually innervated dendritic spines develop in the absence of excitatory activity and resist plasticity through tonic inhibitory crosstalk. *Neuron* **2023,** *111* 3, 362-371. e6.
(43) Errico, V.; De Ninno, A.; Bertani, F. R.; Businaro, L.; Bisegna, P.; Caselli, F., Mitigating positional dependence in coplanar electrode Coulter-type microfluidic devices. *Sensors and Actuators B: Chemical* **2017,** *247*, 580-586.
(44) Bradski, G., The opencv library. *Dr. Dobb's Journal: Software Tools for the Professional Programmer* **2000,** *25* 11, 120-123.
(45) Harris, C. R.; Millman, K. J.; Van Der Walt, S. J.; Gommers, R.; Virtanen, P.; Cournapeau, D.; Wieser, E.; Taylor, J.; Berg, S.; Smith, N. J., Array programming with NumPy. *Nature* **2020,** *585* 7825, 357-362.
(46) McKinney, W., pandas: a foundational Python library for data analysis and statistics. *Python for high performance and scientific computing* **2011,** *14* 9, 1-9.
(47) Van der Walt, S.; Schönberger, J. L.; Nunez-Iglesias, J.; Boulogne, F.; Warner, J. D.; Yager, N.; Gouillart, E.; Yu, T., scikit-image: image processing in Python. *PeerJ* **2014,** *2*, e453.
(48) Hunter, J. D., Matplotlib: A 2D graphics environment. *Computing in science & engineering* **2007,** *9* 03, 90-95.
(49) Allan, D. B.; Caswell, T.; Keim, N. C.; van der Wel, C. M.; Verweij, R. W., soft-matter/trackpy: v0. 6.1. *Zenodo* **2023**.
(50) Stauffer, C.; Grimson, W. E. L. In *Adaptive background mixture models for real-time tracking*, Proceedings. 1999 IEEE computer society conference on computer vision and pattern recognition (Cat. No PR00149), IEEE: 1999; pp 246-252.
(51) Rudin, L. I.; Osher, S.; Fatemi, E., Nonlinear total variation based noise removal algorithms. *Physica D: nonlinear phenomena* **1992,** *60* 1-4, 259-268.
(52) Canny, J., A computational approach to edge detection. *IEEE Transactions on pattern analysis and machine intelligence* **1986,** 6, 679-698.
(53) Rosal, R., Morphological description of microplastic particles for environmental fate studies. *Marine Pollution Bulletin* **2021,** *171*, 112716.
(54) Crocker, J. C.; Grier, D. G., Methods of digital video microscopy for colloidal studies. *Journal of colloid and interface science* **1996,** *179* 1, 298-310.
(55) Virtanen, P.; Gommers, R.; Oliphant, T. E.; Haberland, M.; Reddy, T.; Cournapeau, D.; Burovski, E.; Peterson, P.; Weckesser, W.; Bright, J., SciPy 1.0: fundamental algorithms for scientific computing in Python. *Nature methods* **2020,** *17* 3, 261-272.
(56) Pedregosa, F.; Varoquaux, G.; Gramfort, A.; Michel, V.; Thirion, B.; Grisel, O.; Blondel, M.; Prettenhofer, P.; Weiss, R.; Dubourg, V., Scikit-learn: Machine learning in Python. *the Journal of machine Learning research* **2011,** *12*, 2825-2830.
(57) Cheadle, C.; Vawter, M. P.; Freed, W. J.; Becker, K. G., Analysis of microarray data using Z score transformation. *The Journal of molecular diagnostics* **2003,** *5* 2, 73-81.
(58) Biau, G.; Scornet, E., A random forest guided tour. *Test* **2016,** *25* 2, 197-227.
(59) Borup, D.; Christensen, B. J.; Mühlbach, N. S.; Nielsen, M. S., Targeting predictors in random forest regression. *International Journal of Forecasting* **2023,** *39* 2, 841-868.
(60) Chen, P.; Albert, B. J.; Gao, C.; Alaniva, N.; Price, L. E.; Scott, F. J.; Saliba, E. P.; Sesti, E. L.; Judge, P. T.; Fisher, E. W., Magic angle spinning spheres. *Science advances* **2018,** *4* 9, eaau1540.
(61) Georgiev, R. N.; Toscano, S. O.; Uspal, W. E.; Bet, B.; Samin, S.; van Roij, R.; Eral, H. B., Universal motion of mirror-symmetric microparticles in confined Stokes flow. *Proceedings of the National Academy of Sciences* **2020,** *117* 36, 21865-21872.
(62) Jeffery, G. B., The motion of ellipsoidal particles immersed in a viscous fluid. *Proceedings of the Royal Society of London. Series A, Containing papers of a mathematical and physical character* **1922,** *102* 715, 161-179.
(63) Saffman, P., On the motion of small spheroidal particles in a viscous liquid. *Journal of Fluid Mechanics* **1956,** *1* 5, 540-553.
(64) Altman, L. E.; Hollingsworth, A. D.; Grier, D. G., Anomalous tumbling of colloidal ellipsoids in Poiseuille flows. *Physical Review E* **2023,** *108* 3, 034609.
(65) Cai, X.; Li, X.; Bian, X., Dynamics of an elliptical cylinder in confined Poiseuille flow. *Physics of Fluids* **2024,** *36* 8.
(66) Huang, H.; Lu, X.-Y., An ellipsoidal particle in tube Poiseuille flow. *Journal of Fluid Mechanics* **2017,** *822*, 664-688.
(67) Goda, K.; Ayazi, A.; Gossett, D. R.; Sadasivam, J.; Lonappan, C. K.; Sollier, E.; Fard, A. M.; Hur, S. C.; Adam, J.; Murray, C., High-throughput single-microparticle imaging flow analyzer. *Proceedings of the National Academy of Sciences* **2012,** *109* 29, 11630-11635.
(68) Akselrud, C. I. A., Random forest regression models in ecology: Accounting for messy biological data and producing predictions with uncertainty. *Fisheries Research* **2024,** *280*, 107161.